\documentclass[preprint,journal=jpcafh]{achemso}

\usepackage{graphicx} 
\usepackage{amsmath}
\usepackage{multirow}
\usepackage{hyperref}
\usepackage{cleveref}
\usepackage{mhchem}
\usepackage{natmove}
\usepackage{xcolor}
\usepackage{epstopdf}
\usepackage{lineno}  
\usepackage{gensymb}
\usepackage{subcaption}

\AppendGraphicsExtensions{.tif}

\colorlet{review}{black}

\title{On the existence of distinct equilibrium configurations under orienting external electric fields}

\author{Duc Anh Lai}
\author{Devin A. Matthews}
\email{damatthews@smu.edu}
\affiliation{Department of Chemistry, Southern Methodist University, Dallas, TX 75275, USA}

\begin{document}

\maketitle

\begin{tocentry}
\includegraphics[width=8cm]{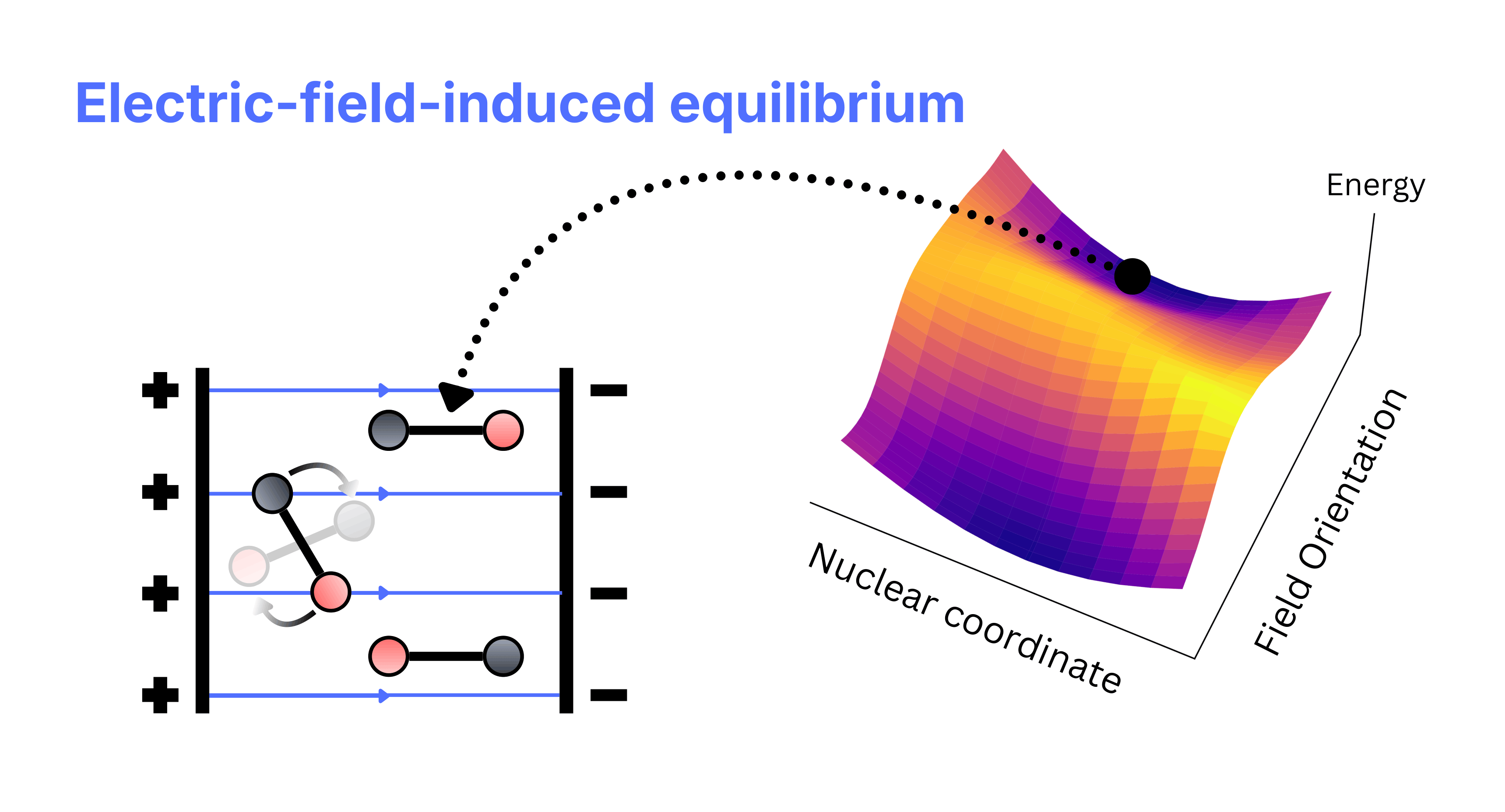}
\end{tocentry}

\begin{abstract}
Oriented external electric fields 
\textcolor{review}{(OEEFs)} 
are ubiquitous in chemistry; however, the effects of fields applied in different directions on molecular systems remain underexplored. A major challenge is that an applied field exerts a torque on a molecule, reorienting the molecular frame and complicating the interpretation of orientation-dependent electric-field effects. 
\textcolor{review}{
Thus, free polar molecules experience orienting rather than oriented fields, such that the field response drives the molecular orientation rather than the other way around.}
\textcolor{review}{In this work, we explore a new regime of distinct molecular equilibrium configurations in the presence of OEEFs, differing in the relative direction of the external field and the molecular frame. While static dipole moments typically dominate molecular responses to external electric fields, these equilibria are enabled by exploiting molecular polarizability.}
These distinct ``directomers'' exhibit unique electronic and nuclear configurations, particularly in their low-lying excited states. We employ oriented electric field vectors referenced to a molecule-fixed principal axis frame along with hybrid analytical-numerical geometry optimization in order to explore the rotational potential energy surface (rRES), as well as a simply analytic model based on equilibrium electrical properties which captures the double-well character of the rPES, including some geometry relaxation effects.

\end{abstract}

\section{Introduction}

Within the scope of the Born-Oppenheimer approximation, the energy landscape in which total energy is calculated from each nuclear configuration is referred to as the potential energy surface (PES) \cite{truhlar_potential_2003}. In chemistry, exploration of PES reveals the underlying mechanism of various physicochemical behaviors, such as molecular stability, reactivity, dynamics, and the capability of undergoing chemical transformations.\footnote{We do not consider thermal contributions to free energy which are of course critical for consideration of finite temperature reactivity. These contributions are relatively minor for the rotational potential energy surfaces discussed here.} Based on the intrinsic properties of a chemical system, potential (free) energy surfaces may exhibit multiple energy minima representing unique conformations, isomers, tautomers, or reaction intermediates, as well as transition states (either singly or as generalized transition surfaces), conical intersection seams, and avoided crossings which connect different minima on surface as well as distinct surfaces together \cite{truhlar_current_1983, domcke_conical_2011}. These features drive a wide range of phenomena such as dynamic stability, isomerization pathways, reaction selectivity, bifurcation, quantum delocalization, and quantum tunneling \cite{osborn_reaction_2017, ess_bifurcations_2008, slavicek_preference_2001, li_ab_2023}. The shape and features of the PES can be controlled to some extent by external parameters, such as solvent, addition of catalysts, electric currents, external fields, and others. More recently, double-well PESs have become a key research direction in the field of switchable systems. The idea behind switchable agents lies in distinctive minimum energy states that allow the system to interconvert under an external stimulation. By controlling the conversion of equilibrium states, one has the capability to regulate internal properties of a chemical system making it possible to design controlled switching devices \cite{yang_high_2023}, molecular motors \cite{feringa_art_2017}, thermal energy or data storage materials \cite{han_electric-field-driven_2020}, biological agents \cite{feringa_art_2017}, and dynamic catalysts \cite{feringa_art_2017}. 

Over the past few years, there has been a rising interest in electric field-assisted chemistry, leading to various notable applications of this research line \cite{shaik_electric-field_2020, shaik_oriented_2025, siddiqui_designed_2023, yang_utilization_2025, eberhart_methods_2025}. An external electric field perturbs both nuclear and electronic structure of atoms and molecules, resulting in energy shifts in molecular spectra \cite{shaik_structure_2018}, modification of molecular properties \cite{stuyver_external_2020, sowlati-hashjin_chemical_2013}, and conformational conversion \cite{alemani_electric_2006, foroutan-nejad_dipolar_2016}. External electric fields, particularly oriented external electric fields (OEEFS), have also been proven to affect various chemical reactions, such as electrostatic reactions \cite{wang_oriented_2019}, Diels-Alder reactions \cite{shaik_electric-field_2020, aragones_electrostatic_2016}, electrophilic aromatic substitution reactions \cite{stuyver_electrophilic_2019}, front-side Menshutkin reactions \cite{ramanan_catalysis_2018}, Woodman--Hoffmann-conflicted pericyclic reactions \cite{shaik_structure_2018}, phase transition \cite{eddy_electric_2024}, charge transfer \cite{jankowska_electric_2015, fruchtl_electronic_2023}, and methyl-transfer reactions  \cite{dutta_dubey_solvent_2020}. Furthermore, the applications of electric fields were reported in the catalytic model of cytochrome P450 complexes \cite{shaik_oriented_2016, lai_external_2010}, biological probe design \cite{suydam_electric_2006, zheng_two-directional_2022}, and protein engineering \cite{hekstra_electric-field-stimulated_2016, martin_electric_2018, sowlati-hashjin_electrostatic_2021}. 

Electric fields are ubiquitous in chemistry and play important roles in many chemical phenomena. Materials like ferroelectrics have intrinsic charge separation that normally creates a homogeneous electric field upon its microscopic surface \cite{wang_electric-field-induced_2025, chen_enabling_2023}, or can be engineered by an external field \cite{negi_ferroelectric_2023}. Additionally, the oriented electric field induced by the resting potential of a bilayer membrane plays a crucial role in cell transportation of biomolecules embedded in those systems \cite{ermakov_electric_2023, zhang_interplay_2019}. Macromolecules containing various ions and charged regions, such as proteins, lipids, and nucleic acids naturally exhibit polarization that generates internal electric fields in their surrounding environment or catalytic cavities \cite{eberhart_methods_2025}, which is the fundamental idea of vibrational Stark spectroscopy \cite{fried_measuring_2015}, and stimulates biological mechanics \cite{hekstra_electric-field-stimulated_2016}. These specifically-directed electric fields are implicated in various phenomena, for example DNA replication, enzyme catalysis, signal transduction, protein folding, fluorescence protein quantum yield, and protein-ligand interactions \cite{drobizhev_local_2021, lehle_probing_2005, ray_effects_2024, bim_local_2021, laberge_intrinsic_1998, hekstra_electric-field-stimulated_2016}. A number of pharmaceutical formulations and active therapies, such as electroporation, iontophoresis, transcutaneous electrical nerve stimulation (TENS), and tissue-engineered muscles (TEMs), harvest an external electric field pulse to regulate actions of drugs or responses of target cells resulting improvements in clinical efficacy and selectivity \cite{mirvakili_wireless_2021, nuccitelli_role_2003, ho_functional_2014, pullar_physiology_2016}. 
\textcolor{review}{Furthermore, many artificial electric-field-based systems inherently impose geometric constraints on molecular orientation. Examples include designed catalytic active sites, single-molecule electronic junctions, surface-adsorbed molecular systems, crystalline environments, and metal-organic framework cavities, where molecular orientation is mechanically or chemically restricted with respect to an applied electric field \cite{zheng_two-directional_2022, yang_utilization_2025, welbornComputationalOptimizationElectric2018}. In such systems, the relative orientation between the molecular framework and the local electric field is dictated by the surrounding environment rather than by free molecular rotation. Rational design of these field-generating motifs has consequently emerged as a powerful strategy for tuning chemical reactivity, catalytic selectivity, charge transport, and molecular properties \cite{eberhart_methods_2025, leonard_electric_2021}. Similarly, molecules in condensed phases or molecular assemblies generally do not adopt a single preferred orientation, and can experience very strong \emph{in situ} electric fields, for example fields up to 50--90 MV/cm${}^{-1}$ (0.01--0.018 a.u.) in oil--water interfaces\cite{shi_water_2025}. Intermolecular interactions, steric constraints, and local electrostatic environments induce heterogeneous molecular orientations, causing individual molecules to experience local electric fields from different directions. Consequently, the local electric field is often not aligned with the permanent molecular dipole moment, making the exploration of equilibrium structures and electronic states under non-parallel oriented external electric fields both physically relevant and practically important.}
Therefore, understanding effects of OEEFs,
\textcolor{review}{especially non-aligned electric fields,}
is particularly vital to elucidate chemical behaviors in practical situations. 

Nevertheless, most electric field-based studies currently focus on the effects of strongly orienting fields that strictly align molecular systems based on the stabilizing interaction of the external field and the permanent molecular dipole moment, in part due to challenges in generating and characterizing non-parallel external fields. 
\textcolor{review}{Experimentally, controlling the relative orientation between a molecule and an applied electric field remains highly challenging \cite{shaik_oriented_2016}. Likewise, many computational studies simplify the problem by aligning the electric field with the permanent dipole moment and implicitly supposing that the field does not rotate the molecular principal axis frame during structural relaxation \cite{rinconActivationSbondsElectric2016, shaik_structure_2018, ramanan_catalysis_2018, wang_oriented_2019, stuyver_electrophilic_2019}. As a result, to the best of our knowledge, the molecular response to non-parallel fields has not been thoroughly investigated. Moreover, measurements performed on molecular ensembles generally reflect a convolution of molecules experiencing different local field orientations, making the underlying microscopic response difficult to interpret. The present work addresses this gap by systematically investigating molecular equilibrium structures and electronic states under arbitrary electric-field orientations, thereby providing a theoretical framework for isolating and understanding the effects of non-parallel external electric fields.}

\textcolor{review}{Previously, the interaction between the molecular dipole moment and the applied electric field has served as the primary approximation for describing molecular responses to electric fields \cite{shaik_structure_2018, yang_utilization_2025, longImpactElectricFields2025, shaik_electric-field_2020, shaik_oriented_2025, stuyver_external_2020, siddiqui_designed_2023, bofillControllingChemicalReactivity2022}.}
However, the polarizability and hyperpolarizability of anisotropic systems induces unique chemical responses to various external field orientations which can modulate or dominate the external field response. In addition, non-alignment of the external field and dipole moment can induce distinctive behavior. For example, heteronuclear molecules exposed to an OEEF exhibit dual properties based on the relative orientation between the field vector and the bond axis \cite{shaik_structure_2018}. Such ``exotic'' electronic configurations are not easily observable, however, due to the extreme destabilization cause by opposing field and dipole moment vectors.
\textcolor{review}{Nevertheless, they become physically relevant in the aforementioned environments, where molecular reorientation is inhibited or where sufficiently strong intrinsic fields exist to orient or bias molecules to directomeric states.}

In this work, we postulate and explore an OEEF interaction regime which provides access to distinct, stable molecular orientations which differ in the relative direction of the external field and permanent dipole moment, which is enabled by leveraging the anisotropic polarizability of asymmetric but 
\textcolor{review}{low-polarity}
molecules. Such ``directomers'' display distinct nuclear and electronic configurations, and can be characterized by a double-well rotational potential energy surface (rPES) with respect to the applied external field angle.
In order to account for non-rigidity of the nuclear framework, we measure the applied field orientation with respect to the molecule-fixed rigid body (inertial) frame. A hybrid geometry optimization approach which utilizes both constrained analytic gradient and numerical line-search schemes is employed to overcome the unavailability of the fully gradient-based optimization. 
\textcolor{review}{Two prototypical low-polarity molecules, carbon monoxide (CO) and carbonyl sulfide (OCS), which possess linear equilibrium structures in the absence of an external electric field, are investigated under an oriented field strength ranging from 0.01 a.u. to 0.05 a.u.}
Both the ground and selected excited states are investigated, with low-lying valence-to-Rydberg states dominating the 
\textcolor{review}{UV-Vis}
spectrum at high field strengths. A simple analytical model based on the equilibrium (field-free) static electrical properties is also developed which reproduces the essential double-well behavior of the rPES as well as the bulk of the geometric relaxation effect.

\section{Theoretical Methods}

\subsection{Field-Dependent Energy Expansion}

\begin{figure}[t]
    \centering
    \includegraphics[width=0.8\linewidth,]{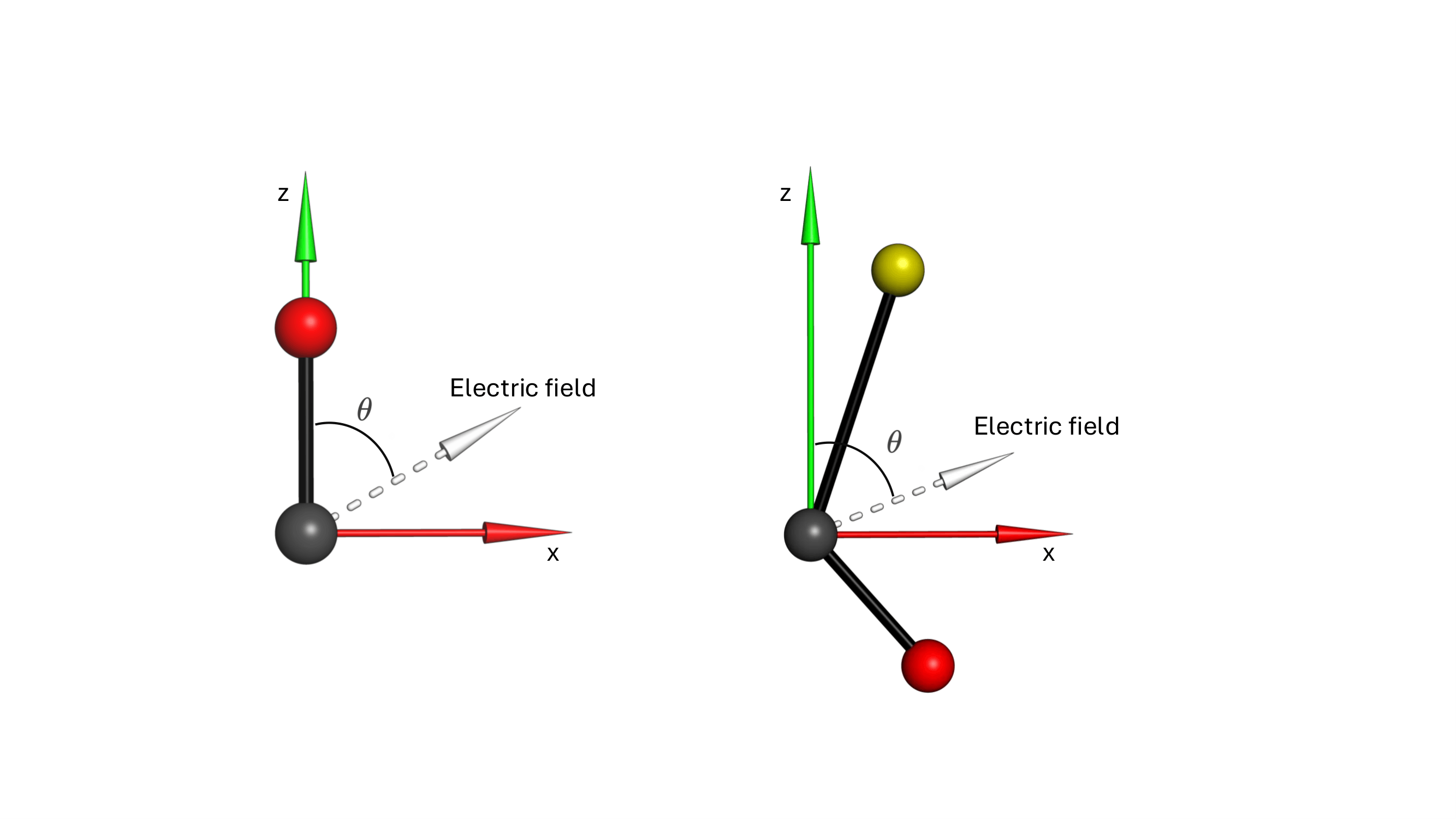}
    \caption{\color{review}{Molecular framework of CO (left) and OCS (right). The geometries are fixed on the $xz$ plane. The electric field vector (in white) is determined by a clockwise rotation from the $z$ axis by an angle $\theta$. The carbon, oxygen, and sulfur atoms are represented in gray, red, and yellow, respectively. For \ce{OCS}, the $z$ axis is aligned with the ``$a$'' principal inertial axis which coincides with the molecular axis at the linear geometry.}}
    \label{fig:mol_axes}
\end{figure}

In this study, we focus on investigating the rotational PES with respect to the field orientation. 
\textcolor{review}{Within this framework, the electric field direction is described in the molecular principal axis frame, which is defined by the three mutually orthogonal eigenvectors of the moment of inertia matrix. Herein, we will reorient the molecules so that the principal axis frame is aligned with the Cartesian coordinate system (\Cref{fig:mol_axes}).}
The advantage of utilizing the principal axis system is that it is invariant to the molecular rotation and is directly related to the equilibrium geometry of the molecule, which is parametrically dependent on the application of the external field. Consequently, the principal axis frame simplifies the determination of the relative orientation between the applied oriented field and the molecular system. Since the molecules considered in this study are linear (planar under applied non-collinear fields), we will always assume that the static electric fields are applied along the $xz$ plane which defines the molecular framework. The field strength can then be decomposed into $x$- and $z$-components which are functions of the angle $\theta$ between the applied electric field and the molecular principal axis which is maximally aligned with the molecular axis, 
\begin{align}
     \varepsilon_x&=\sin\theta\,\Vert\varepsilon\Vert \\
     \varepsilon_z&=\cos\theta\,\Vert\varepsilon\Vert
\end{align}
It is important to note that we define $x$- and $z$-directions according to the frame of the moment of inertia that is the same as regular Cartesian coordinate in the case of linear geometries ($C_{\infty v}$ symmetry) but different in the case of bending geometries ($C_s$ symmetry). During a geometry optimization process, the field direction is constantly referenced at a fixed angle $\theta$ with respect to the \textcolor{review}{``$a$''} principal axis which is the eigenvector associated with the \textcolor{review}{smallest} eigenvalue of the moment of inertia tensor, \textcolor{review}{ corresponding to the interatomic axis in the linear geometry}.

The energy of a linear molecule under a static electric field can be expressed as a function of the field orientation $\theta$,
\begin{align} \label{eq:full series}
    E(\theta,\Vert\varepsilon\Vert)=E_0-\mu_z\cos\theta\Vert\varepsilon\Vert-\frac12\alpha_{xx}\sin^2\theta\Vert\varepsilon\Vert^2-\frac12\alpha_{zz}\cos^2\theta\Vert\varepsilon\Vert^2-\ldots
\end{align}where the zeroth order term ($E_0$) is the value of field-free energy ($\varepsilon=0$), $\boldsymbol\mu$ and $\boldsymbol\alpha$ are the permanent dipole moment vector and polarizability tensor (matrix), respectively. From \Cref{eq:full series}, the perturbed energy 
\textcolor{review}{is a  nonlinear function of both the electric field strength and its orientation relative to the molecular frame.}
\textcolor{review}{In practice, high-order terms in the energy expression are often neglected, and the dipole approximation is typically sufficient to describe molecular responses to external electric fields \cite{shaik_structure_2018, yang_utilization_2025, siddiqui_designed_2023, stuyver_external_2020, longImpactElectricFields2025, shaik_oriented_2025}. For example, Hoffmann and co-workers demonstrated a linear correlation between field-induced perturbation in activation barriers and reaction energies in the weak-field regime \cite{hoffmann_linear_2022}. This linearity, however, breaks down at higher field strengths, where contributions from molecular polarizability become significant \cite{zhaoExploringLinearEnergy2025}. Likewise, for molecules which exhibit substantial polarizability and/or hyperpolarizability (specifically anisotropy of these quantities, vide infra), higher-order responses can alter the rotational PES, giving rise to additional stationary points and distinct equilibrium orientations.}

\subsection{Dipole-Polarizability model}

\textcolor{review}{To describe the nonlinear response of a molecule under an external electric field, we propose using the second-order truncated formula of \Cref{eq:full series},
\begin{align} \label{eq:truncated}
    E(\theta,\Vert\varepsilon\Vert)=E_0-\mu_z\cos\theta\Vert\varepsilon\Vert-\frac12\alpha_{xx}\sin^2\theta\Vert\varepsilon\Vert^2-\frac12\alpha_{zz}\cos^2\theta\Vert\varepsilon\Vert^2
\end{align} which captures the dipole- and polarizability-dependent responses to the applied field. One can use  \Cref{eq:truncated} to approximate the electric-field-perturbed potential energy surface over the parameter space defined by the field orientation ($\theta$) and the field strength ($\Vert\varepsilon\Vert$). Based on the quadratic form, modeling the rotational potential surface using the dipole-polarizability model enables the rPES to transition from a single-well to a double-well shape.}
In the typical dipole-dominated regime, either $0\degree$ or $180\degree$ is a minimum and the other a maximum, depending on the directionality of the permanent dipole moment. However, under sufficiently strong fields and/or suitable electronic properties both points may become minima, connected by a maximum at $0\degree<\theta<180\degree$.\footnote{These points are minima assuming the typical case $\Delta\alpha<0$. If the molecule is more polarizable along the perpendicular direction then these become maxima and the rPES has a single minimum at a non-collinear field orientation.} This situation occurs whenever the following condition is satisfied,
\begin{align} \label{eq:condition}
    -\Vert\varepsilon\Vert<\frac{\mu_z}{\Delta\alpha}< \Vert\varepsilon\Vert
\end{align}
As might be expected, this mathematical constraint involves the relationship between the total field strength $\Vert\varepsilon\Vert$, the static dipole moment along the principal moment of inertia $\mu_z$, and the polarizability anisotropy $\Delta\alpha=\alpha_{xx}-\alpha_{zz}$. While the dipole moment plays a significant role in the response of a molecule to an external electric field, the anisotropic polarizability determines the majority of the distinctive molecular response to different orientations of the external field.

From \Cref{eq:condition}, typical chemical systems with a large dipole moment and relatively small polarizability will not exhibit two minima as well as a potential energy barrier, but a monotonic curve from 0 to $180^\circ$ of the field direction. However, intentional choice of systems with these desired properties can lead to alternative rPES landscapes under applied fields. Moreover, the condition infers the field direction $\theta_\text{max}$ that becomes the quasi-transition state and determines whether the \textcolor{review}{activation barrier on the} PES is skewed to the left or to the right of the field angle range. The model also predicts approximate potential energies corresponding to stationary points as,
\begin{align}
    &\text{At }\theta=0^\circ:&\quad E_1&=E_0-\mu_z\Vert\varepsilon\Vert-\frac{1}{2}\alpha_{zz}\Vert\varepsilon\Vert^2 \\
    &\text{At }\theta=180^\circ:&E_2&=E_0+\mu_z\Vert\varepsilon\Vert-\frac{1}{2}\alpha_{zz}\Vert\varepsilon\Vert^2 \\
    &\text{At }\theta_{\max}=\arccos\frac{\mu_z}{\Delta\alpha\Vert\varepsilon\Vert}:&E^\ddagger&=E_0-\frac{\mu_z^2}{2\Delta\alpha}-\frac{1}{2}\alpha_{xx}\Vert\varepsilon\Vert^2
\end{align}
The forward and backward energy barriers are then computed as,
\begin{align}
    \Delta E^\ddagger_1&
=\mu_z\Vert\varepsilon\Vert-\frac{1}{2}\Delta\alpha\Vert\varepsilon\Vert^2 -\frac{1}{2}\frac{\mu_z^2}{\Delta\alpha} \\
    \Delta E^\ddagger_2
&=-\mu_z\Vert\varepsilon\Vert-\frac{1}{2}\Delta\alpha\Vert\varepsilon\Vert^2 -\frac{1}{2}\frac{\mu_z^2}{\Delta\alpha}
\end{align}
The energy difference between two minima linearly depends only on the molecular dipole moment and the total field strength,
\begin{align}
    \Delta E=2\mu_z\Vert\varepsilon\Vert
\end{align}

\subsection{Hybrid optimization scheme}

Under the application of a non-parallel external field, linear polyatomic molecules such as OCS may exhibit bending motion, and hence break the symmetry from  linear ($C_{\infty v}$ for OCS) to planar ($C_s$). In the inertial frame, this bond-angle relaxation will rotate the applied field vector and introduce a new term in the nuclear gradients. To tackle this problem, we implement a hybrid optimization scheme that separates each geometry relaxation into two stages: the internal (analytical) geometry optimizations are utilized to relax bond lengths at a constrained bond angle, whereas the bond angle relaxation is handled externally using a line-search algorithm. The line-search algorithm starts with a slightly bent geometry at $179.9^\circ$ and follows the bisection algorithm to locate the bond angle range that contains the minimum energy. 
Having obtained a set of discrete energies corresponding to different bond angles, the line-search optimizer fits a cubic spline function to interpolate the next step for the bond angle. 
\textcolor{review}{The principal axes are recalculated for each geometric update (for both bond-length and bond-angle relaxations), such that the electric field vector remains referenced to the molecular framework. While changes in bond length in each optimization iteration do also affect the orientation of the moments of inertia, and hence the external field, the effect is relatively small compared to the bond-angle changes and does not adversely affect the analytical geometry optimization. Co-iteration of the bond-length and bond-angle optimization stages ensures convergence to the true minimum geometry at the specified field angle.}
Convergence criteria is satisfied when either the change in bond angle is below $10^{-3}$ degrees or the change in bond length and energy are less than $10^{-6}$ and $10^{-10}$ a.u., respectively. 

\subsection{Computational details}

The bond-length optimization is performed using the Q-Chem 6.0 package \cite{epifanovskySoftwareFrontiersQuantum2021}. The first- and second-order electrical properties corresponding to the static dipole moment and molecular polarizability, respectively, were analytically calculated using the CFOUR program suite \cite{matthewsCoupledclusterTechniquesComputational2020}. The natural bond orbital analysis were performed using the NBO version 5.0 \cite{glendening_natural_2012}. All calculations are performed at the frozen core coupled cluster with single and double excitations (CCSD)\cite{purvis_full_1982} level of theory with the aug-cc-pVTZ basis set,\cite{kendall_electron_1992} utilizing the equation-of-motion coupled cluster method with single and double excitations (EOM-CCSD)\cite{sekino_linear_1984,geertsen_equation--motion_1989,comeau_equation--motion_1993,stanton_equation_1993} for obtaining excitation energies.

\section{Results}
\subsection{Rotational potential energy surface of a diatomic molecule: CO}

\begin{figure}[t]
    \centering
    \includegraphics[width=\linewidth,]{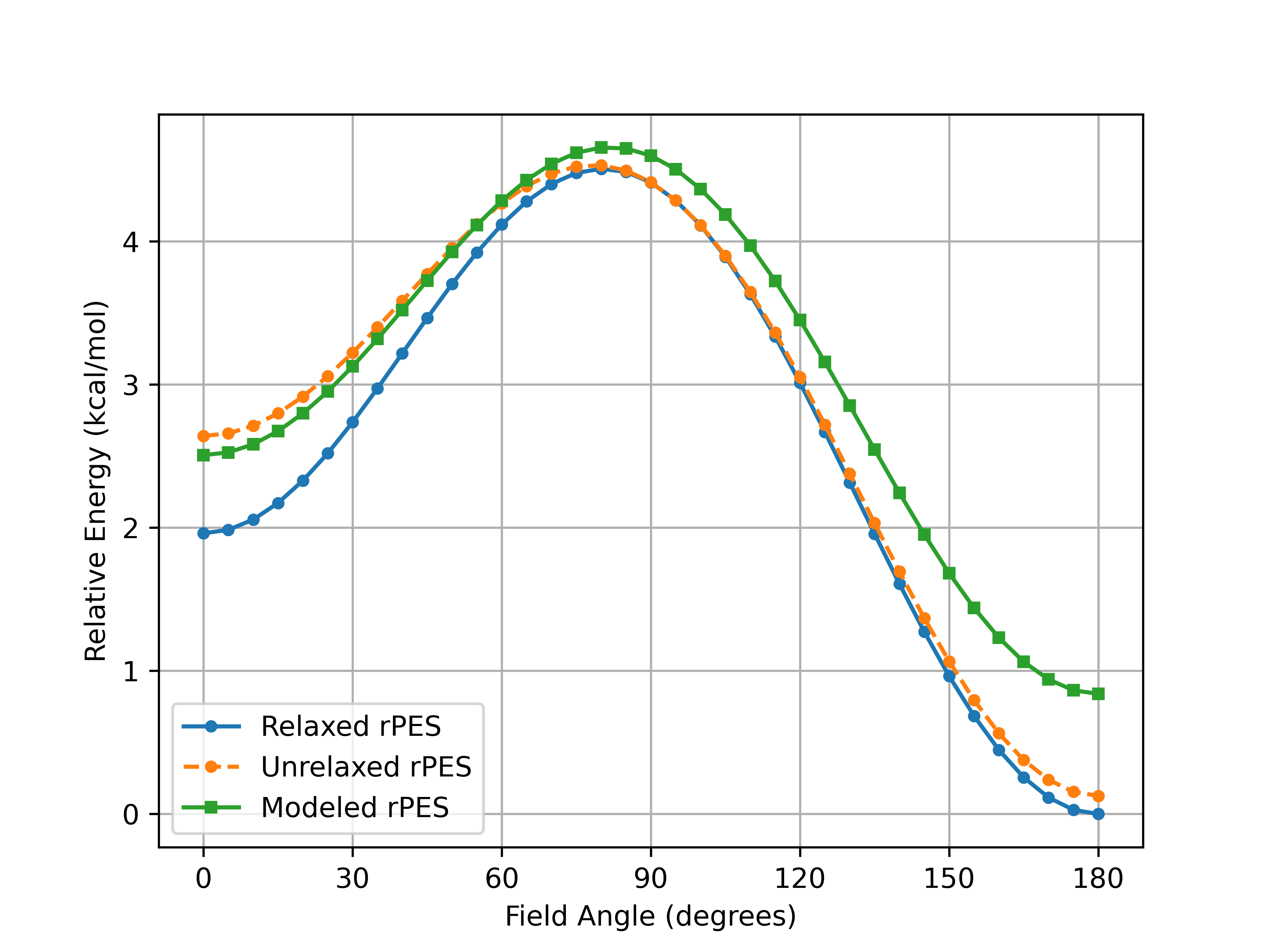}
    \caption{Rotational potential energy surface of the ground state of CO at a strength 0.05 a.u.}
    \label{fig:pes_gr-CO}
\end{figure}

First, we examine effects of an OEEF on the CO molecule by exploring the rPES of the ground state and the first excited state
\textcolor{review}{in the Franck--Condon ordering at field strengths ranging from 0.01 a.u. to 0.05 a.u.}
The molecule is oriented along the Cartesian $z$-axis, in which C$\rightarrow$O indicates the positive direction. The ground state rPES under OEEFs from 0.01 a.u. to 0.05 a.u. exhibits two minima at 0 and $180^\circ$ of the field angle. A ($-z$)-field parallel to the permanent dipole moment (O$\rightarrow$C), 
\textcolor{review}{induces the greatest degree of polarization (Table S1), resulting in a global minimum on the rPES.}
The off-bond-axis fields have a lesser impact on the 
\textcolor{review}{field-induced stabilization, primarily through the reduction in polarizability in this direction compared to along the molecular axis (vide infra), thereby}
giving rise to a barrier on the rPES \textcolor{review}{(\Cref{fig:pes_gr-CO})}. However, the quasi-transition state emerges at a field angle somewhat less than $90^\circ$ ($80.4^\circ$), indicating the energy is not only governed by the dipole, but also the anisotropy of the polarizability contributions. Although having \textcolor{review}{an insignificant effect} on the
\textcolor{review}{dipole moment component along the}
bond axis, the perpendicular field generates an induced dipole moment orthogonal to the bond to the largest extent. For instance, $\mu_x=1.533$ Debye at a field strength of 0.05 a.u. (Table S1). Aligning the field towards the C$\rightarrow$O direction,
\textcolor{review}{which is anti-parallel to the permanent dipole moment,}
the field induces depolarization along the bond axis, and subsequently reverses the molecular dipole moment. Since the field strength is sufficient to reverse dipole moment which likewise stabilizes the molecule, a local minimum energy appears at the parallel ($+z$)-external field. In this case, the external field needs to counterbalance the static dipole moment before flipping the molecular polarization direction, e.g. the amplitude of dipole moment at $180^\circ$ field is 2.25 Debye compared with 2.04 Debye at $0^\circ$ field under the field strength 0.05 a.u. As a result, the minimum energy at $0^\circ$ is higher than that at $180^\circ$ \textcolor{review}{(\Cref{tab:CO-fields})}. In addition, increasing electric field strength leads to larger differences between the two minima and between the two minima and the maximum (\Cref{tab:CO-fields}). The unique responses to parallel and anti-parallel fields of heteronuclear diatomic molecules are also reported by other ab initio studies \cite{foroutan-nejad_dipolar_2016, sowlati-hashjin_chemical_2013, shaik_impact_2021}, although only for parallel $(+z)$ and $(-z)$ fields. Considering the change of bond length, when orienting the field from 0 to $180^\circ$ the \ce{C-O} distance gradually decreases, from 1.152 \AA\ to 1.119 \AA\ under the field strength of 0.05 a.u. (Table S1). 
\textcolor{review}{Given that the field-free \ce{C-O} bond length is 1.129 \AA,} the maximum compression of the bond (at $180\degree$) is approximately half as much as the maximum lengthening (at $0\degree$).

\begin{table}[t]
    \centering
    \begin{tabular*}{\textwidth}{@{\extracolsep{\fill}} llllll @{}}
    \hline
        Field strength (a.u.) & 0.01 & 0.02 & 0.03 & 0.04 & 0.05 \\ \hline
        $\theta_{\max}$ (deg.) & 50.09 & 70.63 & 76.58 & 79.10 & 80.36 \\
        $\Delta E$ (kcal/mol) & 0.3355 & 0.6832 & 1.0571 & 1.4749 & 1.9599 \\
        $\Delta E^{\ddagger}_1$ (kcal/mol) & 0.0171 & 0.2416 & 0.7291 & 1.4895 & 2.5458 \\
        $\Delta E^{\ddagger}_2$ (kcal/mol) & 0.3526 & 0.9248 & 1.7862 & 2.9644 & 4.5057 \\
        \hline
    \end{tabular*}
    \caption{Ground-state rotational PES of CO molecule under various field strengths. $\theta_{\max}$: dihedral angle at energy maximum; $\Delta E$: energy difference between two minima; $\Delta E^{\ddagger}_1$: energy difference between maximum and minimum (at $0^\circ$ field); $\Delta E^{\ddagger}_2$: energy difference between maximum and minimum (at $180^\circ$ field).}
    \label{tab:CO-fields}
\end{table}

\begin{figure}
    \centering
    \includegraphics[width=\linewidth]{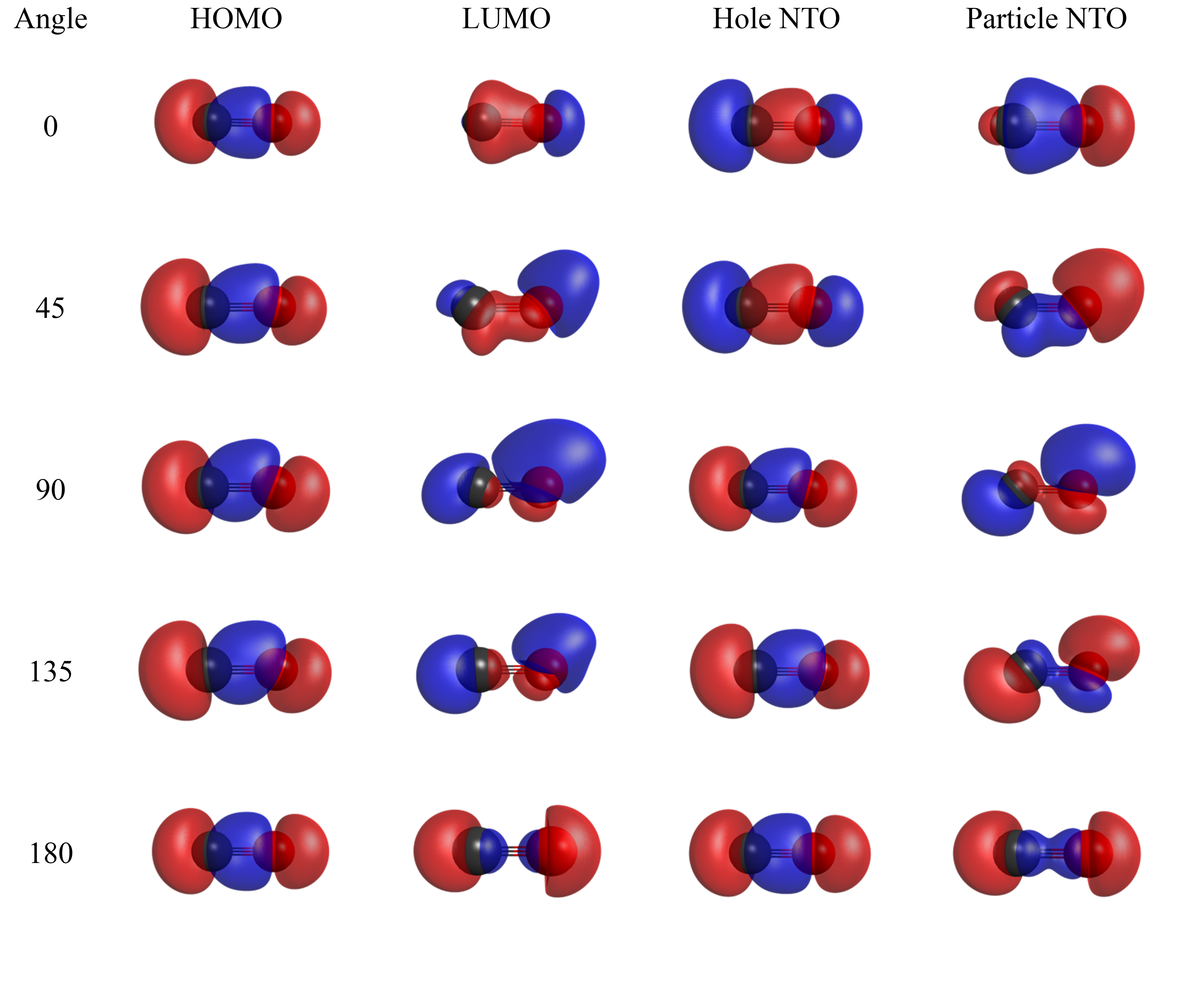}
    \caption{Iso-density surfaces of HOMO, LUMO, and particle/hole natural transition orbital (NTO) pair for the CO ground and first excited state under a field strength of 0.05 a.u. HOMO and hole NTO surfaces are plotted with an isovalue of 0.05, while the LUMO and particle NTO surfaces are plotted with an isovalue of 0.01.}
    \label{fig:co_orbitals}
\end{figure}

To form a deeper understanding of the changes to electronic structure under oriented external fields, we analyzed the natural bond orbitals of the ground state of the CO molecule at a field strength of 0.05 a.u. In the field-free ground state the molecule exhibits one $\sigma$ bond and two $\pi$ bonds, with strict separation based on the corresponding symmetry species. 
\textcolor{review}{The natural bond orbital for the $\sigma$-bond closely resembles the HOMO shown in \Cref{fig:co_orbitals}, with subtle changes in local orientation at the atomic centers defining the natural hybrid orbital (NHO) angles as discussed below. These changes are not as readily evident from visual inspection compared to the the virtual orbitals.}
Applying electric fields along the molecular axis unsurprisingly retains the orthogonality between the $\sigma$ and $\pi$ orbitals. The molecular orbitals at $0^\circ$ and $180^\circ$ show a conventional $\sigma$ bond formed between the C and O atoms. In contrast, switching the electric field to varying angles between these extremes, the loss of axial symmetry collapses the orthogonality between the $\sigma$ and $\pi_x$ orbitals (although $\pi_y$ remains orthogonal as it transforms as $A''$ rather than $A'$ as $\sigma/\pi_x$), enabling the $\sigma$ orbital to hybridize with $\pi_x$.
This effect of symmetry collapse and forbidden orbital mixing is a hallmark of OEEFs\cite{shaik_impact_2021}. Previously, Li and co-workers \cite{li_anisotropic_2020} demonstrated the anisotropic Stark effect induced by a perpendicular electric field that allows $\sigma$ orbitals to incorporate $\pi$ character. Our results generalize this observation for an arbitrary angle of the field and highlights the changing nature of $\sigma$--$\pi$ hybridization along the rPES. The increased mixing of the (both doubly occupied) $\sigma$ and $\pi$ orbitals at field angles near 90\degree\ also rationalizes the energy barrier between the parallel and anti-parallel minima.
Natural hybrid orbital (NHO) directional analysis reveals that both hybrid orbitals of carbon and oxygen atoms are bent away from the bond axis under the exposure of an off-bond-axis field (Table S1). This phenomenon might be a key factor in understanding the changes observed in the system under the influence of the external field.

The unrelaxed rPES computed from the field-free equilibrium geometry is used to compare with the relaxed curve. As shown in \Cref{fig:pes_gr-CO}, unrelaxed field stabilization energies are consistently underestimated. The quasi-energy minimum is still preserved at $0^\circ$, but with a moderate error of $0.680$ kcal/mol. The principal reason for this difference is that the unrelaxed electron density favors a reversed polarization structure of the molecule in the parallel field. In order to create a bond dipole of reverse orientation, the \ce{C-O} bond must significantly lengthen. On the other hand, orienting the field towards the anti-parallel direction of the molecular principal axis favors the intrinsic dipole moment, and hence results in smaller errors ($0.126$ kcal/mol at $\theta=180\degree$). Given that the amount of 
\textcolor{review}{bond}
compression is only half of the amount of bond extension, the quadratic nature of the potential energy curve (PEC, that is energy w.r.t. $R_{\ce{C-O}}$) aligns well with a roughly $5\times$ reduction in error due to neglect of bond relaxation.

In the absence of the field, the dipole moment, $xx$-polarizability, and $zz$-polarizability of the CO molecule are analytically computed as $-0.0266$ a.u. ($-0.0626$~D), $11.7138$ a.u., and $15.4427$ a.u., respectively. Inserting these electric properties into the second-order expansion 
\textcolor{review}{(\Cref{eq:truncated}),} we obtain a modeled rPES. 
\textcolor{review}{An advantage of this modeling approach
is that it requires minimal computational effort, as only field-free energies and electric properties are needed. Consequently, it is well suited for rapid assessments of electric-field-induced potential energy surfaces.}
Despite the simplification, the quadratic expression is capable of retaining the character of the true rPES. In terms of the energy difference between minima, the second-order function improves on the unrelaxed curve, but underestimates the energy barrier and overall stabilization compared to the relaxed rPES. The energy errors vary between $0.14$ and $0.84$ kcal/mol, where the maximum error occurs at $180^\circ$. The prediction of the transition state angle is $81.8^\circ$ which is only $1.4^\circ$ in error from the relaxed value. Although, the model always predicts higher values for potential energy than the actual ones, we discover that the predicted forward and backward barrier are $2.15$ kcal/mol and $3.82$ kcal/mol, which differ by $0.40$ kcal/mol and $0.69$ kcal/mol from the relaxed values, respectively. Comparison to higher-level calculations of CO polarizabilities and hyperpolarizabilities\cite{doi:10.1021/jp960412n} suggests that errors due to underestimation of the polarizability total less than 0.25 kcal/mol, while lack of hyperpolarizability terms accounts for approximately 0.6 kcal/mol at 0 and 180\degree. In the $(+z)$ direction, \textcolor{review}{the modeled and unrelaxed rPESs agree well; because the unrelaxed rPES implicitly includes hyperpolarizability contributions through application of a finite field, but does not include any correction for geometric relaxation, this similarity implies that the hyperpolarizability error in this direction is minimal. On the other hand, in the ($-z$) direction, the unrelaxed and relaxed curves nearly coincide, indicating that relaxation effects are much less important in this configuration. Indeed, the \ce{C-O} bond is significantly less compressed at $180\degree$ than it is extended at $0\degree$ (Figure~S1). The modeled rPES experiences a similar error at $0\degree$ and $180\degree$, which due to the fact that the only significant remaining error at $180\degree$ is through neglect of hyperpolarizability, that geometric relaxation and hyperpolarizability errors are of overall similar magnitude but distributed very differently along the rPES.}

\begin{figure}[t]
    \centering
    \begin{subfigure}[b]{0.49\textwidth}
        \includegraphics[width=\textwidth]{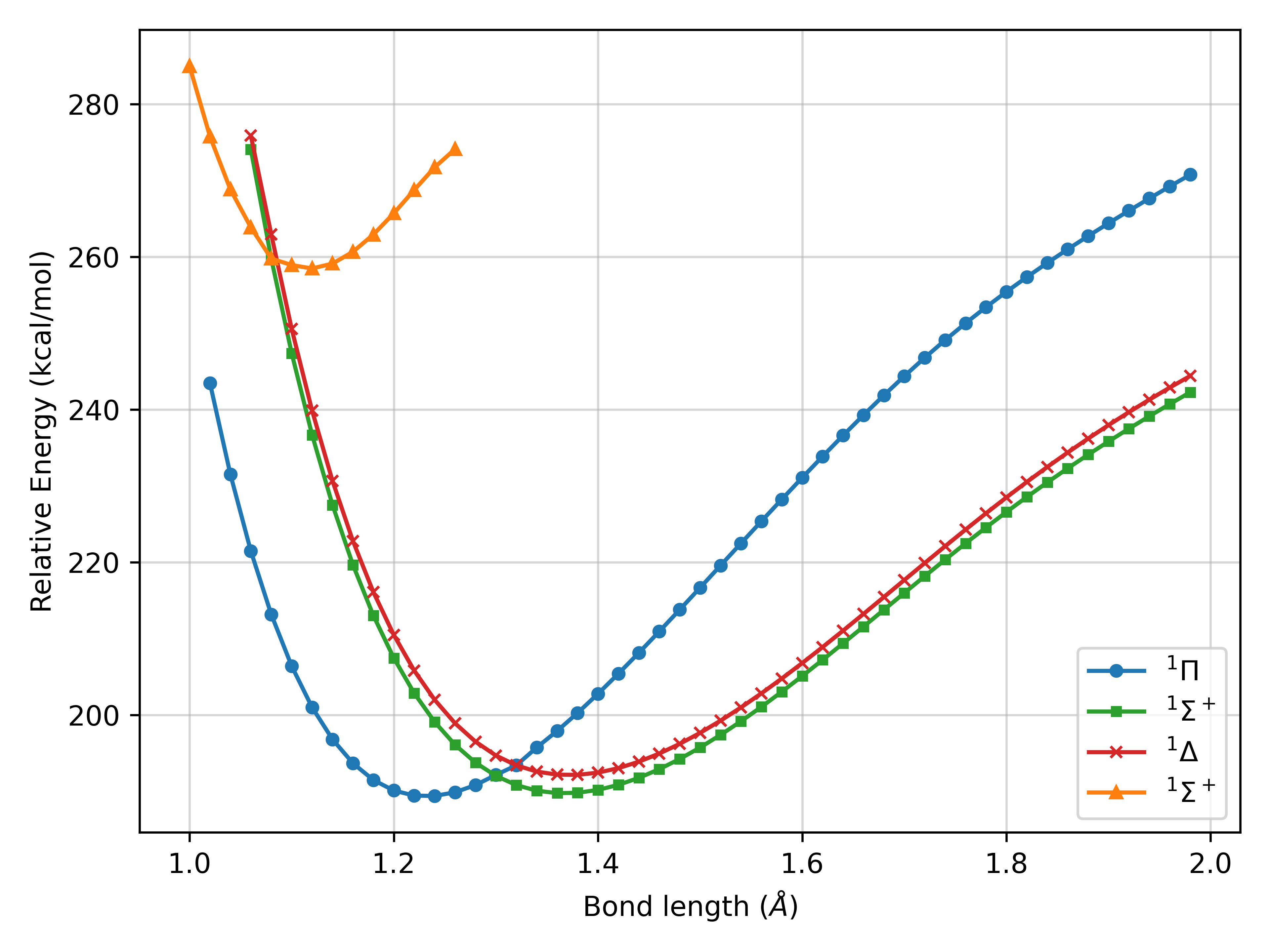}
        \caption{Field-free}
        \label{fig:scan_nofield}
    \end{subfigure}
    \begin{subfigure}[b]{0.49\textwidth}
        \includegraphics[width=\textwidth]{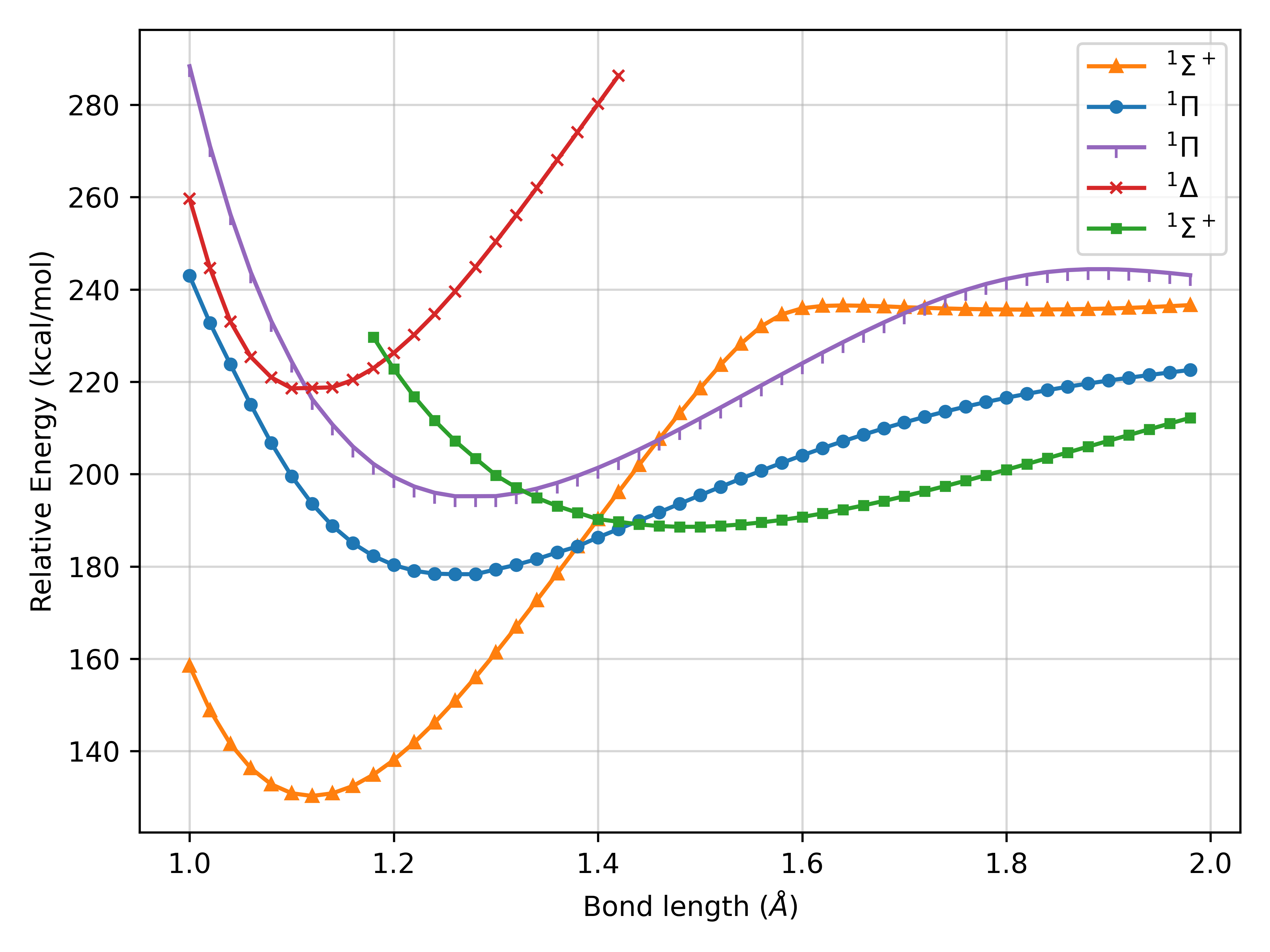}
        \caption{0.05 a.u. field applied along C$\rightarrow$O}
        \label{fig:scan_field}
    \end{subfigure}
    \caption{Potential energy curves of low-lying excited-states of the CO molecule with and without an applied external field.} %
\end{figure}

\begin{figure}[t]
    \centering
    \includegraphics[width=\linewidth]{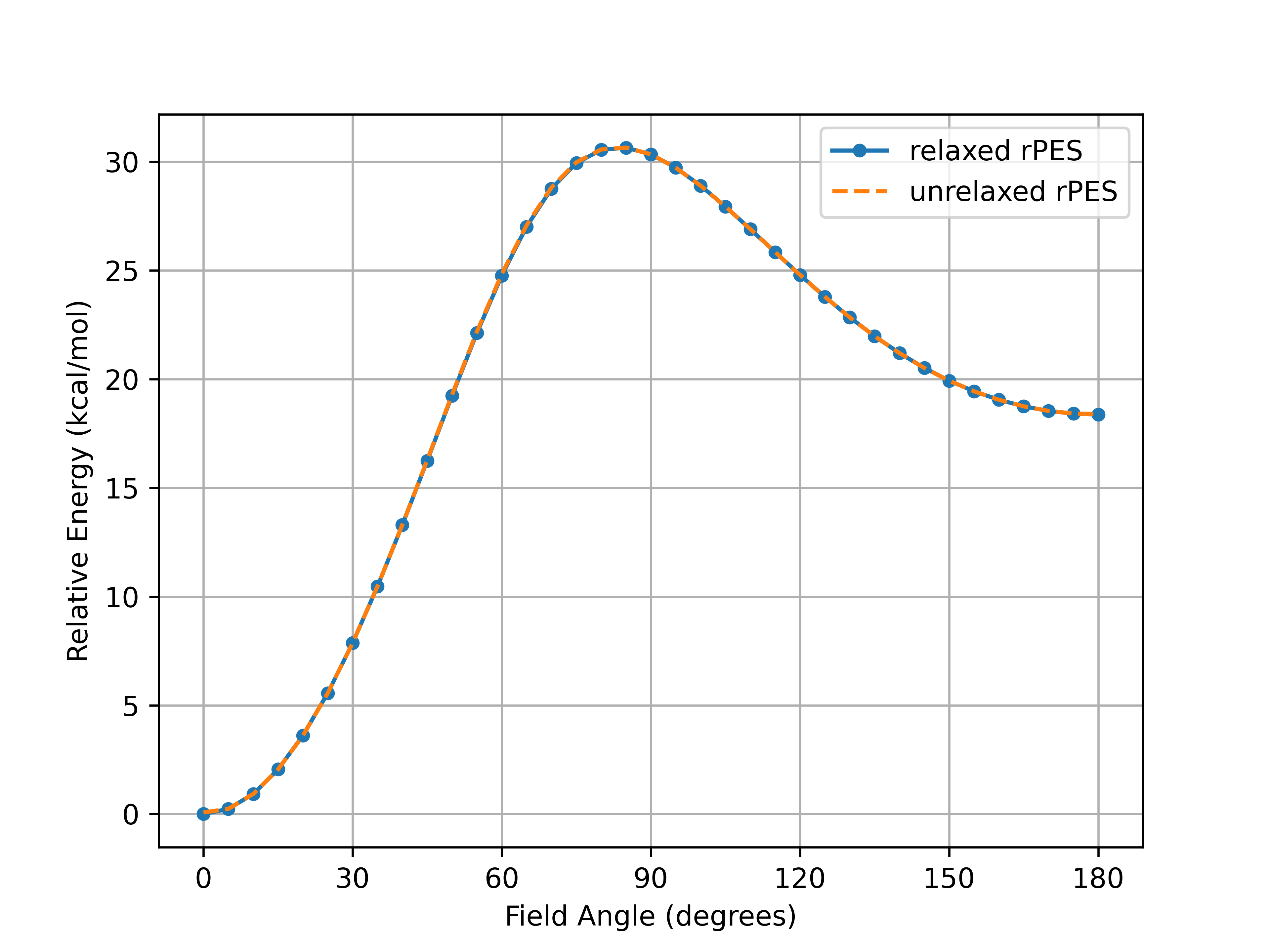}
    \caption{Rotational potential energy surface of the first excited state of CO at a strength 0.05 a.u.}
    \label{fig:pes_ex-CO}
\end{figure}

The effects of external electric fields are significant on electronically excited transitions. In the absence of fields, the lowest excited states of CO are found to have dominant $\sigma\rightarrow\pi^*$ (${}^1\Pi$) or $\pi\rightarrow\pi^*$ (${}^1\Sigma^+$/${}^1\Delta$) character (\Cref{fig:scan_nofield}). However, electric fields ranging from 0.01 a.u. to 0.05 a.u. significantly stabilize the $\sigma\rightarrow \text{Ry(3s)}$ (upper ${}^1\Sigma^+$) state, and it becomes the lowest energy excited state in the presence of sufficiently strong applied fields (see \Cref{fig:scan_field}). The first excited state rPES of CO under a field strength of 0.05 a.u. is shown in \Cref{fig:pes_ex-CO}. Two energy minima are still located at $0^\circ$ and $180^\circ$ field angles. However, the energy minimum at parallel ($+z$) field is lower than that at anti-parallel ($-z$) field, indicating a non-trivial increase in the excited state dipole moment ($\mu_z\approx -0.7$~D) compared to the ground state. The quasi-maximum occurs when the field is oriented at $83.6^\circ$. The forward and backward barriers are significantly higher than in the ground state at 30.7 kcal/mol and 12.3 kcal/mol, respectively. As shown in Table~S2, the relaxed \ce{C-O} bond length in the first excited state under various field orientations is remarkably similar to the field-free equilibrium value of 1.118 \AA. As a result, the unrelaxed rPES of the first excited-state shows minimal difference compared with the relaxed rPES (\Cref{fig:pes_ex-CO}).

We also computed isodensity surfaces for the canonical HOMO and LUMO orbitals, as well as the particle and hole natural transition orbitals (NTOs) of the lowest excited state, as depicted in \Cref{fig:co_orbitals}. 
\textcolor{review}{Typically, valence anti-bonding virtual orbitals (e.g., $\pi^*,\sigma^*$) are well represented by canonical Hartree-Fock or Kohn-Sham virtual orbitals, thus dominant hole and particle NTOs often resemble the corresponding canonical frontier orbitals in valence excitations.  Surprisingly, under the application of an external electric field, the particle NTO of the Rydberg first excited state in \ce{CO} also matches closely with the LUMO despite the delocalized (spatially as well as in the canonical virtual basis) nature of Rydberg orbitals, as well as potential entanglement of multiple NTOs. This indicates a highly Koopman-like nature of the excitation despite the expected diffuse character and orbital mixing associated with Rydberg states.}
\textcolor{review}{The atomic $p$ orbital contributions to the particle NTO are no longer aligned with the nodal plane strictly containing the bond axis. Their orientations (qualitatively determined by a angular deviation from the nodal plane near the atomic center) are tilted from the field-free orientation even moreso than in the computed ground state NHO angles (Table S1), and up to almost perpendicular to the bond axis, e.g. particle NTO at 90\degree-field-angle (\Cref{fig:co_orbitals}).}
Here, the C and O local \textcolor{review}{orbital} orientations do not necessarily co-rotate, as the particle NTO (and to a lesser extent the HOMO) near the O atom is rotated in the same direction \textcolor{review}{relative to the \ce{C-O} bond axis} for all $\theta$ between 0\degree\ and 180\degree, while near C the direction of rotation \textcolor{review}{reverses}
between 45\degree\ and 90\degree. External electric fields, especially those oriented non-collinear to the permanent dipole moment can thus be seen to impact the virtual orbitals and hence excited state electronic structure significantly moreso than in the ground state. As with the excited state rPES (\Cref{fig:pes_ex-CO}) in to comparison with the ground state, this difference is due to the much higher polarizability via the spatially diffuse character of the virtual orbital space, although the electric field response may not be more anisotropic in a relative sense.

\subsection{Rotational potential energy surface of a triatomic molecule: OCS}

In this section we present the behavior of the OCS molecule, as a model triatomic model, within an oriented electric environment of 0.05 a.u. field strength. We place the O$\rightarrow$C$\rightarrow$S direction along the positive $z$-axis. Unlike in CO, OCS has three independent internal coordinates which suggest a more diverse and potentially unpredictable response of the molecule to the OEEF. The ground state of OCS in the absence of an external field exhibits a linear geometry where the equilibrium \ce{C-O} and \ce{C-S} bond lengths are 1.15 \AA\ and 1.57 \AA, respectively. 

\begin{figure}[t]
    \centering
    \includegraphics[width=0.9\linewidth]{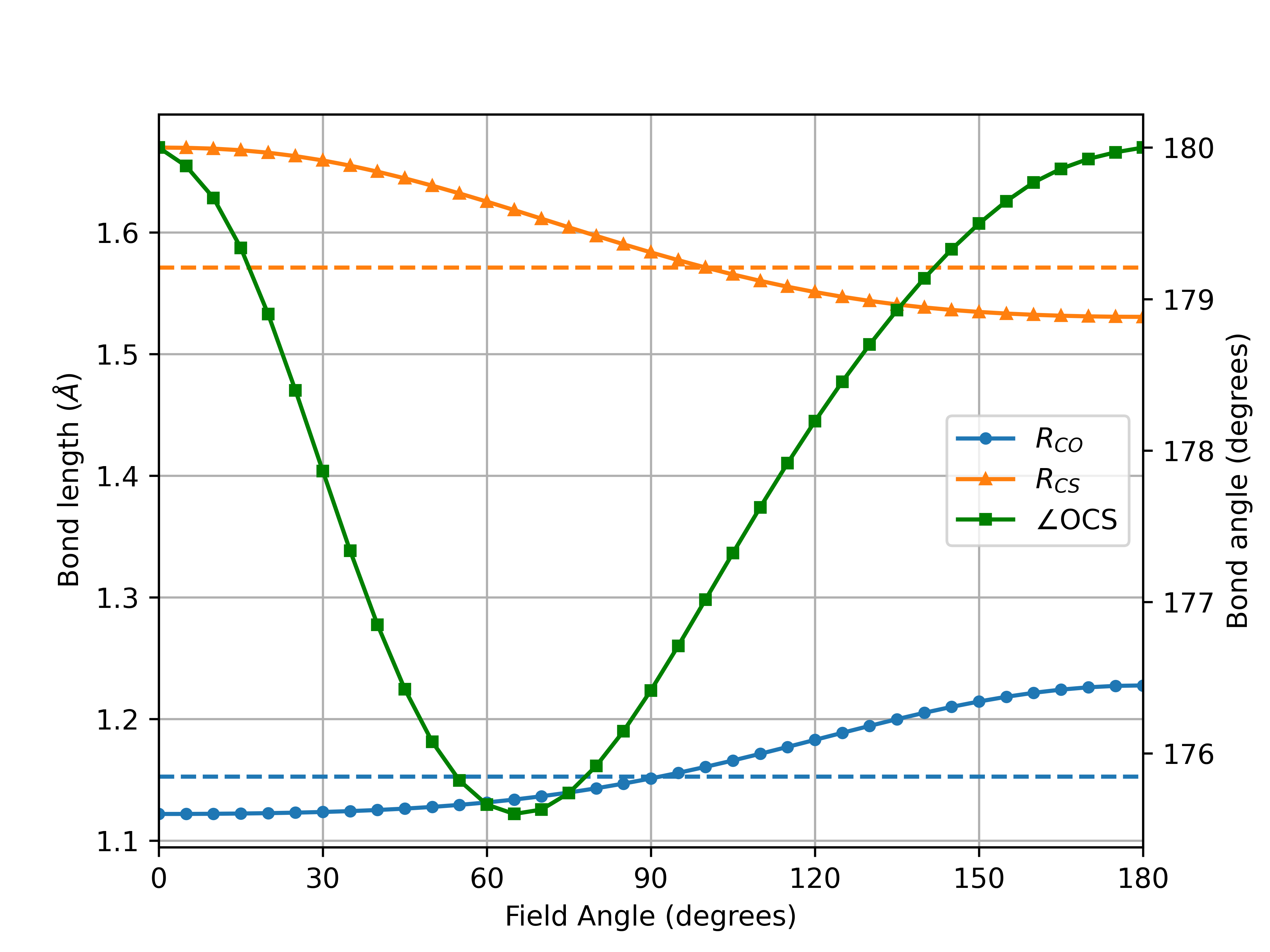}
    \caption{Structural parameters of OCS under various field orientations and field strength of 0.05 a.u. Dashed lines indicate equilibrium bond lengths in the field-free ground state.}
    \label{fig:struc_ocs_gr}
\end{figure}


When an OEEF with 0.05 a.u. magnitude is applied, there is 
a inverse 
trend in the relaxation of \ce{C-S} and \ce{C-O} bond lengths (\Cref{fig:struc_ocs_gr})\textcolor{review}{, with one bond lengthening while the other contracts and vice versa}. Rotating the field from 0 to $180^\circ$ slightly lengthens the distance between the carbon and the oxygen atom from 1.12~\AA\ to 1.23~\AA. Simultaneously, the \ce{C-S} bond length steadily decreases from 1.67~\AA\ to 1.53~\AA\ within the same field angle range. Additionally, bending the molecule lowers the total potential energy which obviously reduces the symmetry from linear ($C_{\infty v}$) to the non-linear $C_s$ group. It is interesting to note that the most highly bent geometry does not occur at a precisely perpendicular external field. As shown in \Cref{fig:struc_ocs_gr}, the bond angle asymmetrically decreases from the linear geometries and reaches a minimum of $175.6^\circ$ when the field is oriented towards $65.8^\circ$. The optimizations for field angles of 0\degree\ and 180\degree\ with respect to the principal axis frame were started at slightly non-linear geometries, showing that alignment of the field to the $c$ principal axis recovers a linear geometry naturally.

\begin{figure}[t]
    \centering
    \includegraphics[width=\textwidth,height=\textheight,keepaspectratio]{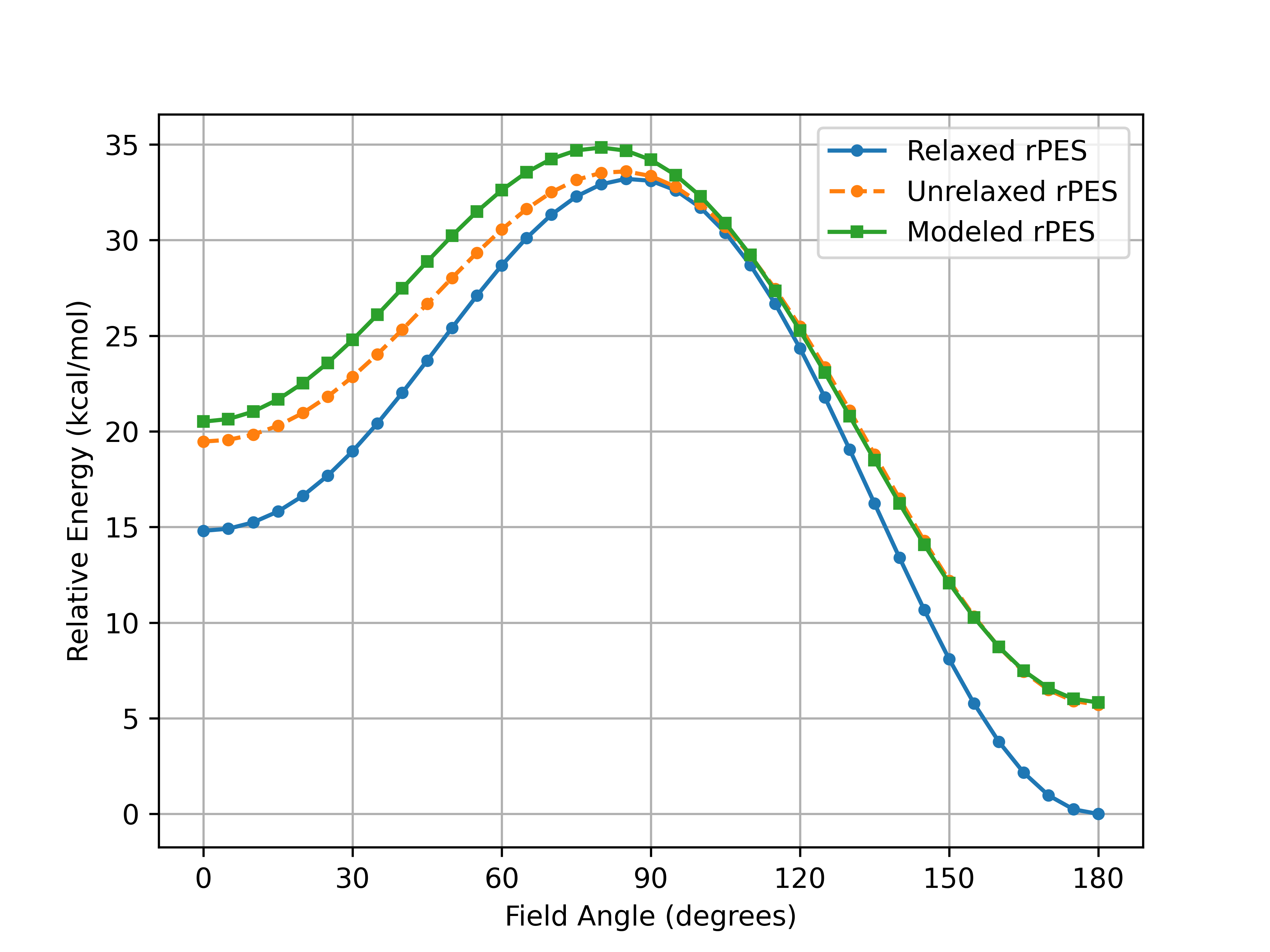}
    \caption{Rotational Potential energy surface of the ground state of OCS molecule under effects of OEEF of strength 0.05 a.u.}
    \label{fig:pes_gr_OCS}
\end{figure}

The field-perturbed ground state rPES of OCS also presents a double-well shape with two energy minima occurring at $0^\circ$ and $180^\circ$ (\Cref{fig:pes_gr_OCS}). 
Additionally, the quasi-transition state along the field angle coordinates emerges when the field is applied at $86.3^\circ$, \textcolor{review}{shifting the energy barrier on the} 
rPES slightly towards \textcolor{review}{the parallel field orientation ($\theta=0\degree$).}
Similar to the observation in CO molecule, the minimum energy at the anti-parallel field is lower than that at parallel field. The forward and backward energy barrier are computed as $18.41$ kcal/mol and $33.22$ kcal/mol, respectively, making it thermally more difficult to switch between two stable states than for CO.

\Cref{fig:pes_gr_OCS} shows a comparison between the relaxed and the unrelaxed rPES of the OCS molecule under the 0.05 a.u. field strength. Similar to the CO molecule, the unrelaxed rPES overestimates both minima at $0^\circ$ and $180^\circ$ by $4.67$ kcal/mol and $5.71$  kcal/mol, respectively. The errors at both minima are more similar than in CO owing to the alternate compression/expansion of the \ce{C-O} and \ce{C-S} bonds. In addition, the unrelaxed rPES incorrectly predicts the quasi-transition state barrier, giving foward and reverse barrier heights too low by $4.27$ kcal/mol and $5.32$ kcal/mol, respectively. The difference between relaxed and unrelaxed rPES increases with the molecular size, and hence accurate geometries are essential to modeling field-dependent chemical behaviors accurately for larger molecules. 

In the case of OCS, the field-free electrical properties are $-0.2340$ a.u. ($-0.5948$~D), $25.5803$ a.u., and $52.3880$ a.u. corresponding to the molecular dipole moment, $xx$-polarizability, and $zz$-polarizability, respectively. The second-order power series results in a slightly better prediction of the shape of the true rPES compared to the unrelaxed rPES. As shown in \Cref{fig:pes_gr_OCS}, the modeled rPES predicts the quasi-transition state at 79.$9^\circ$ field angle, an error of $6.4^\circ$ compared to the relaxed rPES. Errors at the linear geometries are highly similar to the unrelaxed rPES, but errors in the forward and reverse barrier heights are somewhat improved at $4.08$ kcal/mol and $4.21$ kcal/mol, respectively, while the error in $\Delta E$ is reduced from $1.04$ kcal/mol in the unrelaxed rPES to $0.12$ kcal/mol.

\begin{figure}[t]
    \centering
    \includegraphics[width=\linewidth]{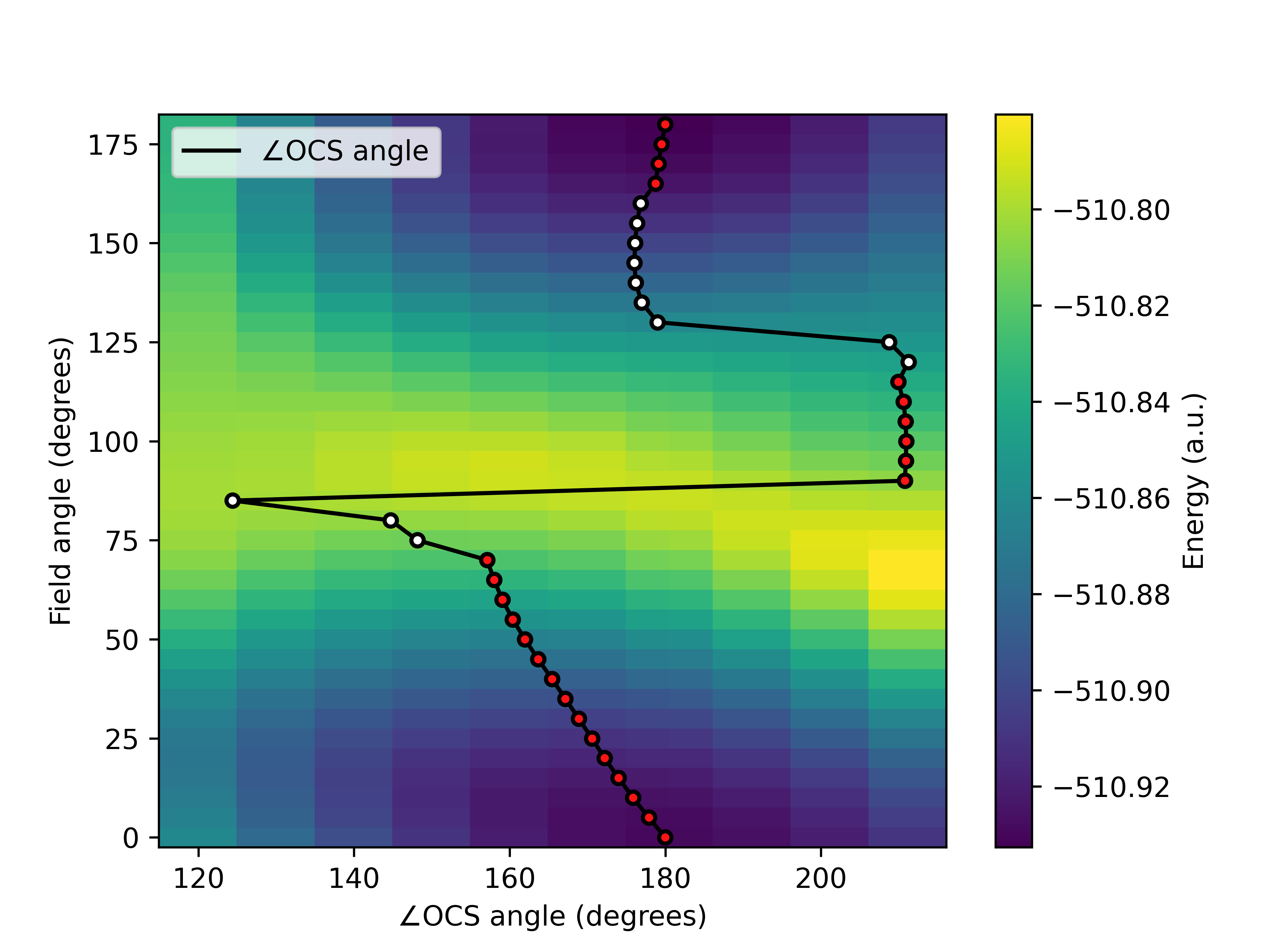}
    \caption{Energy of the globally-lowest excited state of OCS as a function of both external field angle and OCS bond angle, with fully relaxed CO and CS bond lengths. The minimum-energy path from 0\degree\ to 180\degree\ field angle is shown as a solid line, with the minum at each field angle determined via local fitting. White (red) points indicate states of $A'$ ($A''$) symmetry.}
    \label{fig:ocs_scan}
\end{figure}

\begin{figure}[t]
    \centering
    \includegraphics[width=\linewidth]{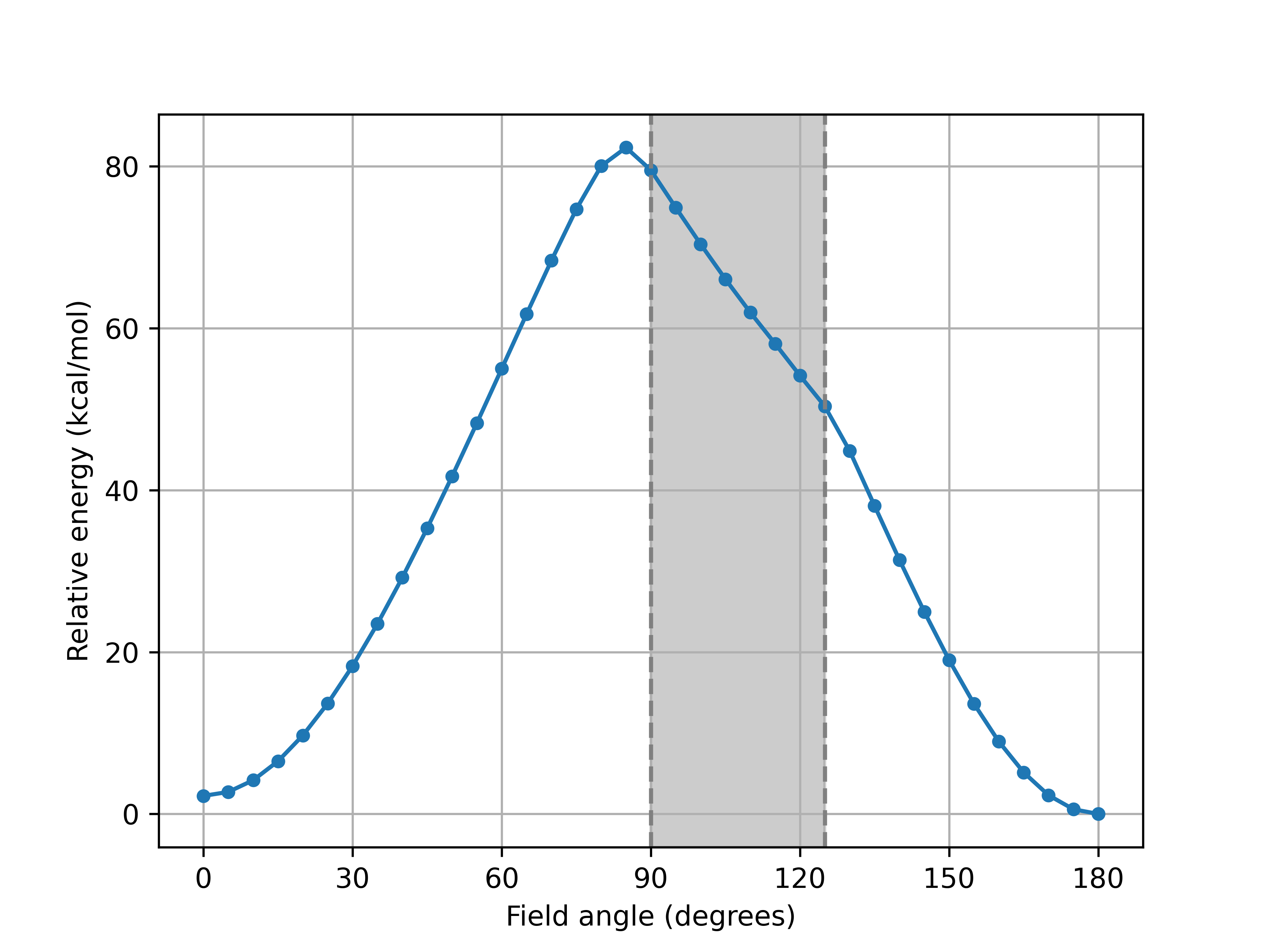}
    \caption{Energy of the lowest excited state, with respect to the ground state, as a function of field angle. Minimum geometries at each field angle are determined as in \Cref{fig:ocs_scan}. The shaded region highlights the significant discontinuities in OCS bond angle at 90\degree\ and 125\degree\ field angles.}
    \label{fig:pes_ex_OCS}
\end{figure}

\begin{figure}[t]
    \centering
    \includegraphics[width=\linewidth]{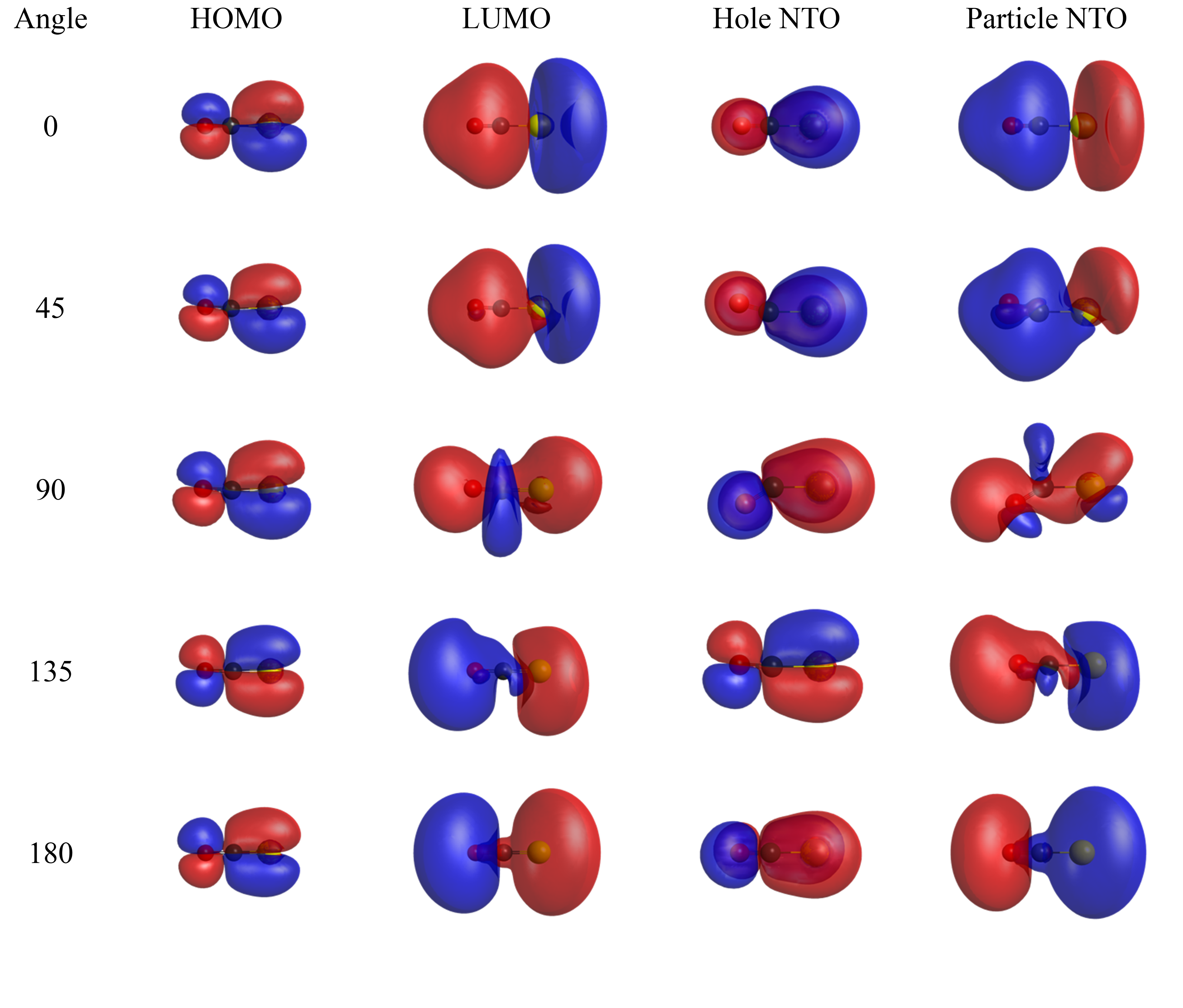}
    \caption{Density isosurfaces of HOMO, LUMO, and particle/hole natural transition orbital (NTO) pair for the OCS ground and first excited state under a field strength of 0.05 a.u. All surfaces are plotted with an isovalue of 0.005.}
    \label{fig:ocs_orbitals}
\end{figure}

The excited state energy landscape of OCS exhibits complex features due to the lowered symmetry, increase in number of degrees of freedom, and congestion of the excited state manifold. Due to the limitations of equation-of-motion coupled cluster theory\cite{10.1063/1.4998724}, we were not able to reliably discriminate between conical intersections and narrowly-avoided crossings, and due to the complexity of the energy landscape we did not attempt to construct full potential energy surfaces (in the field and bond angle subspace) for multiple individual states. Rather, we study the lowest-energy adiabat, regardless of excited state symmetry. This surface is depicted in \Cref{fig:ocs_scan}, as well as a minimum-energy path along the external field angle which was determined by complete global optimization of the OCS first excited state geometry. While the overall double-well shape, further elucidated in \Cref{fig:pes_ex_OCS}, is still present, the simple interpretation in terms of a second-order electrostatic response is lost. Rather, the minimum-energy path exhibits a sharper maximum due to the change of electronic character and symmetry, in particular the discontinuity at 90\degree which implies a radical change in OCS bond angle. This surface exhibits no true (or at least singular) transition state, but the height of the quasi-barrier ($\sim80$ kcal/mol) implies high thermal stability of the linear geometries on this surface. Unlike the ground state in which the energy along bond angles only show a minimum, the excited-state energy surface along the bond angles show a double-well shape when applying non-parallel electric fields (\Cref{fig:ocs_scan}), that is, there are typically distinct local minima for geometries bent both towards and away from the electric field vector. As a consequence, the change in global minimum bond angle for near perpendicular electric fields is discontinuous leading to discontinuous rPES (the gray region in \Cref{fig:pes_ex_OCS}). In addition, the first excited-state rPES of OCS represents two minima at parallel and anti-parallel fields, and they are approximately equal. This observation again confirms that the polarizability plays a key role in energy response. The quasi-transition state exhibits a sharp energy maximum at $84.7^\circ$. 

Orbital density isosurfaces for select field angles are depicted in \Cref{fig:ocs_orbitals}. At both 0\degree\ and 180\degree, the lowest state is of $\pi(\ce{CS})/n(\ce{O})\rightarrow\text{Ry}$ ($\Pi$) character. While the involved occupied orbital (hole NTO) is consistent across the range of angles (with some switching between the $A'$ and $A''$ components which are split by non-parallel fields), the virtual orbital (particle NTO) changes character considerably. While all particle NTOs are highly diffuse (Rydberg type), they are only clearly $ns$ type at the linear geometries, with significant mixing and local reorientation at all non-linear field/molecule geometries. Thus, most of the variation in the electronic character (apart from $A'$/$A''$ switching) is due to Rydberg $s$/$p$ mixing driven by electronic and, to a lesser extent, geometric perturbations. Except at 90\degree, the LUMO and particle NTO show nearly identical character, again confirming that the lowest excitations, while involving transitions to Rydberg-type orbitals, are highly Koopman-like.

\section{Conclusions}

In this work, we investigated the distinct equilibrium molecular and electronic configurations induced by an external electric field applied in different directions relative to the permanent dipole moment, which can be described through the concept of a rotational potential energy surface (rPES). We show that the rPES of two prototypical molecules, CO and OCS, exhibits a double-well profile with two energy minima corresponding to parallel and anti-parallel orientations of the electric field relative to the molecular axis defined in the principal axis frame, giving rise to chemically distinct ``directomers''. Although the unrelaxed rPES computed from field-free equilibrium geometries correctly reproduces the presence of these two minima, the associated energy errors increase with molecular size. Therefore, we suggest that analytic geometry optimization is essential for accurately describing electric-field effects. Furthermore, a second-order Taylor expansion of the energy provides reasonably good agreement with the analytic solutions, highlighting the important role of polarizability in the theoretical description of molecular systems in oriented and orienting external electric fields. While the systems studied here require unphyisically large external fields (0.05~a.u.${}=2.57$~V/\AA), the dipole-polarizability model suggests that searching for additional molecules with large static $\Delta\alpha/\mu$ could uncover systems with more robust directomeric response.

\textcolor{review}{The present framework extends the conventional dipole-aligned description of electric-field effects by treating arbitrary field orientations. Because non-parallel electric fields naturally arise in chemically and mechanically constrained environments--including enzymatic active sites, molecular electronic junctions, surface-adsorbed systems, and porous crystalline materials--our approach provides a theoretical foundation for understanding and predicting field-dependent molecular structure and electronic properties in realistic chemical environments.}

\begin{acknowledgement}
This work was supported in part by the US National Science Foundation (grant CHE-2143725) and by the US Department of Energy (grant DE-SC0022893). Computational
resources for this research were provided by SMU’s O’Donnell Data Science and Research
Computing Institute.
\end{acknowledgement}

\begin{suppinfo}
A Supplemental Information file (.pdf) is available and contains tabular version of the data presented in the article, as well as optimized Cartesian and total dipole moments for each field angle and NHO bond angles for the ground state of CO.
\end{suppinfo}

\bibliography{efield}

\end{document}